\documentclass{aa}
\usepackage{epsfig,times}

\makeatletter
\def\@cite#1#2{(#1\if@tempswa , #2\fi)}
\def\@citex[#1]#2{\if@filesw\immediate\write\@auxout{\string\citation{#2}}\fi
  \def\@citea{}\@cite{\@for\@citeb:=#2\do
    {\@citea\def\@citea{;\penalty\@m\ }\@ifundefined
       {b@\@citeb}{{\bf ?}\@warning
       {Citation `\@citeb' on page \thepage \space undefined}}%
\hbox{\csname b@\@citeb\endcsname}}}{#1}}
\makeatother

\title{GRB010222: afterglow emission from a rapidly decelerating shock
\thanks{Based on observations collected at: the Italian Telescopio
Nazionale Galileo (TNG), operated on the island of La Palma by the Centro
Galileo Galilei of the CNAA (Consorzio Nazionale per l'Astronomia e
l'Astrofisica) at the Spanish Observatorio del Roque de los Muchachos of
the Instituto de Astrofisica de Canarias; the Asiago Astronomical
Observatory, Italy; the Bologna Astronomical Observatory in Loiano, Italy;
the Campo Imperatore Astronomical Observatory, Italy, and; TIRGO
infrared observatory, Switzerland}}

\author{N. Masetti\inst{1}
\and E. Palazzi\inst{1}
\and E. Pian\inst{2,1}
\and F. Mannucci\inst{3}
\and L.A. Antonelli\inst{4}
\and A. Di Paola\inst{4}
\and P. Saracco\inst{5}
\and S. Savaglio\inst{4}
\and L. Amati\inst{1}
\and C. Bartolini\inst{6}
\and S. Bernabei\inst{7,8}
\and D. Bettoni\inst{9}
\and S. Covino\inst{5}
\and S. Cristiani\inst{10}
\and S. Desidera\inst{11}
\and S. Di Serego Alighieri\inst{12}
\and R. Falomo\inst{9}
\and F. Frontera\inst{1,13}
\and F. Ghinassi\inst{14}
\and A. Guarnieri\inst{6}
\and A. Magazz\`u\inst{14}
\and R. Maiolino\inst{12}
\and M. Mignoli\inst{7}
\and L. Nicastro\inst{16}
\and M. Pedani\inst{14} 
\and A. Piccioni\inst{6}
\and B.M. Poggianti\inst{9}
\and V. Testa\inst{4}
\and G. Valentini\inst{15}
\and A. Zacchei\inst{14}
}

\institute{Istituto Tecnologie e Studio delle Radiazioni Extraterrestri,
CNR, Via Gobetti 101, I-40129 Bologna, Italy
\and
Osservatorio Astronomico di Trieste, Via G.B. Tiepolo 11,
I-34131 Trieste, Italy 
\and
Centro per l'Astronomia Infrarossa e lo Studio del Mezzo Interstellare,
CNR, Largo E. Fermi 5, I-50125 Florence, Italy
\and
Osservatorio Astronomico di Roma, via di Frascati 33,
I-00040 Monteporzio Catone, Italy
\and
Osservatorio Astronomico di Brera, via Bianchi 46, I-23807 Merate, Italy
\and
Dipartimento di Astronomia, Universit\`a di Bologna, Via Ranzani 1,
I-40127 Bologna, Italy
\and
Osservatorio Astronomico di Bologna, Via Ranzani 1, I-40127
Bologna, Italy
\and
Instituto de Astrof\'{\i}sica de Canarias, C/ V\'{\i}a L\'actea s/n,
E-38200, La Laguna, Tenerife, Spain
\and
Osservatorio Astronomico di Padova, Vicolo dell'Osservatorio 5,
I-35122 Padua, Italy
\and
Dipartimento di Astronomia, Universit\`a di Padova, Vicolo 
dell'Osservatorio 5, I-35122 Padua, Italy
\and
Osservatorio Astronomico di Asiago, Via dell'Osservatorio 8, I-36012,
Asiago, Italy
\and
Osservatorio Astrofisico di Arcetri, Largo E. Fermi 5, I-50125 Florence,
Italy
\and
Dipartimento di Fisica, Universit\`a di Ferrara, via Paradiso 12, I-44100
Ferrara, Italy
\and
Telescopio Nazionale Galileo, Roque de Los Muchachos Astronomical
Observatory, P.O. Box 565, E-38700 Santa Cruz de La Palma, Spain
\and
Osservatorio Astronomico di Collurania-Teramo, Via Maggini, I-64100
Teramo, Italy
\and
Istituto di Fisica Cosmica ed Applicazioni all'Informatica, CNR, via ugo
La Malfa 153, I-90146 Palermo, Italy
}

\offprints{Nicola Masetti, masetti@tesre.bo.cnr.it}
\date{Received March 19, 2001; Accepted May 23, 2001}

\begin{document}

\abstract{The GRB010222 optical and near-infrared (NIR) afterglow was
monitored at the TNG and other Italian telescopes starting $\sim$1 day
after the high-energy prompt event. The $BVR$ light curves, which are the
best sampled, are continuously steepening and can be described
by two power laws, $f(t) \propto t^{-\alpha}$, of indices 
$\alpha_1 \sim$ 0.7 and $\alpha_2 \sim$ 1.3 before and after a break
occurring at about 0.5 days after the GRB start time, respectively.
This model accounts well also for the flux in the $U$, $I$ and $J$ bands,
which are less well monitored. The temporal break appears to be achromatic.
The two $K$-band points are not consistent with the above behaviour, and
rather suggest a constant trend.
A low-resolution optical spectrum has also been taken with TNG.
In the optical spectrum we found three absorption systems at different
redshifts (0.927, 1.155 and 1.475), the highest of which represents a 
lower limit to, and probably coincides with, the redshift of the GRB.  
The broad-band optical spectral energy distributions do not appear to
vary with time, consistently with the achromatic behaviour of the light
curves. 
We compare our measurements with different afterglow evolution scenarios
and we find that they favor a transition from relativistic to
non-relativistic conditions in the shock propagation. 
}

\maketitle
\markboth{N. Masetti et al.: GRB010222: afterglow emission in a
rapidly decelerating jet}{N. Masetti et al.: GRB010222: afterglow
emission in a rapidly decelerating jet}

\section{Introduction}

The light curves of Optical Transients (OTs) associated with GRBs are
generally described by single power laws $f(t) \propto t^{-\alpha}$ with
indices $\alpha \simeq 1.1-2$. However, in a number of well monitored
cases a change in the light decay rate at about 0.5-1 days after the GRB
was detected with a transition to a steeper power law behaviour.  The
power law indices before and after the temporal break are typically in the
range 0.7-1 and 1.7-2.4, respectively (GRB990123: Fruchter et al. 1999, 
Castro-Tirado et al. 1999, Kulkarni et al. 1999; GRB990510: Stanek et al.
1999, Harrison et al. 1999, Israel et al. 1999; GRB990705: Masetti et
al. 2000a; GRB000926: Fynbo et al. 2001a).
This behaviour can be caused by the  
deceleration of a relativistic jet in a homogeneous medium (Sari et al.
1999, Rhoads 1999), or by expansion in a wind (Chevalier \& Li 1999, 
2000), or by the
transition between relativistic and Newtonian conditions in the plasma
expansion (Dai \& Lu 1999).  Fits to multiwavelength data of individual
afterglows with the above models have been proved to be satisfactory (Dai
\& Lu 1999, Panaitescu \& Kumar 2000).
The multiwavelength optical and near-infrared (NIR) spectra of OTs are
generally well accounted for by synchrotron radiation in a
relativistically expanding shock, and are described by power laws of
different characteristic indices, depending on the location of injection
and cooling breaks, and temporal evolution thereof (Sari et al. 1998).
Both the fading rate and the spectral slopes are determined by
the shape of the electron distribution, in different ways, according to
the geometry of the emitting region.  
Departures of the optical-NIR spectral slopes from those expected based on
the flux temporal decrease are often interpreted as due to absorption of
the OT light within the GRB host galaxy (e.g. Palazzi et al. 1998, Dal
Fiume et al. 2000, Klose et al. 2000, Price et al. 2001a). 

\medskip

GRB010222 was simultaneously detected by the Gamma--Ray Burst
Monitor (GRBM) and the Wide Field Camera (WFC) unit 1 onboard
{\it BeppoSAX} on 2001 Feb 22.30799 UT (Piro 2001a,b) as one of the most
intense GRBs observed by both the GRBM and WFC.
A quick repointing of the satellite allowed the detection of the X--ray
afterglow at a position consistent with that of the prompt event (Gandolfi
2001). A detailed analysis of the high-energy data is presented by in 't
Zand et al. (2001).

Many observers started a ground-based campaign in order to
search for the GRB afterglow at lower wavelengths.
A bright transient source ($R\sim$ 18.4) was independently detected in the
optical by Henden (2001a,b; see also Henden \& Vrba 2001) and McDowell et
al. (2001) about 4 hours after the GRB.
The object, which is not present in the DSS-II sky survey, is at
coordinates (J2000) $\alpha$ =  14$^{\rm h}$ 52$^{\rm m}$ 12$\fs$55,
$\delta$ = +43$^\circ$ 01$'$ 06$\farcs$2 with an error of 0$\farcs$2 along
both directions (Henden \& Vrba 2001), well inside the error box of both
high-energy prompt event and X--ray afterglow as detected by the 
{\it BeppoSAX} Narrow Field Instruments. The OT is among the 
brightest ever observed associated with GRBs.

The fading behaviour of the object, first reported by Henden \& Vrba
(2001) and later confirmed by Stanek et al. (2001a,b), left little 
doubt on its afterglow nature. 
Assuming for the $R$-band flux a power law decay, typical of optical
afterglows, Price et al. (2001b) and Fynbo et al. (2001a) measured a
temporal slope $\alpha$ = 0.89$\pm$0.09 and $\alpha$ = 0.86$\pm$0.01,
respectively. Masetti et al. (2001) reported that a steepening of the
decay slope to $\alpha\sim$ 1.3 occurred at about 1 day after the GRB
start time.

Garnavich et al. (2001) acquired an optical spectrum of the OT soon after
its detection. From the presence of Fe {\sc ii} and Mg {\sc ii} 
absorption features, they measured a redshift $z$ = 1.477 for the
afterglow. Jha et al. (2001a,b) also noticed the presence in the same   
spectrum of a second absorption system, located at $z$ = 1.157. A third
absorption system at $z$ = 0.928 in the optical spectrum
of the afterglow was reported by Bloom et al. (2001). The farthest of the
three systems appeared to be formed by substructures at sligtly different
redshifts, consistent with internal gas motions in galaxies (Castro et
al. 2001).

A bright counterpart was also detected at NIR (Di Paola et al. 2001),
submillimetric (Fich et al. 2001) and radio (Berger \& Frail 2001)
wavelengths.

The good sampling and the optical brightness of the GRB010222 afterglow 
have allowed a detailed study of its evolution at least up to about 40
days after the explosion.
In this paper we present the results of the optical and NIR monitoring
of the optical/NIR transient associated with GRB010222 conducted at 
various Italian telescopes and started about 1 day after
the GRB. Section 2 illustrates the observations and the analysis of the
photometric and spectroscopic data; the results are presented in Section 3
and discussed in Section 4.

\begin{table*}[t!]
\caption[]{Journal of the optical and NIR observations of the GRB010222
afterglow. Magnitude uncertainties are at 1$\sigma$ confidence level}
\begin{center}
\begin{tabular}{rccccl}
\noalign{\smallskip}
\hline
\noalign{\smallskip}
Exposure start & Telescope & Filter & Total exposure &
Seeing & \multicolumn{1}{c}{Magnitude$^1$}\\
\multicolumn{1}{c}{(UT)} & & & time (s) & (arcsec) &  \\
\noalign{\smallskip}
\hline
\noalign{\smallskip}

2001 Feb  22.982 & AZT-24 & $J$ &2700 & 4.0 & 18.6$\pm$0.2$^2$\\
 23.056 & Asiago & $R$ & 1200 & 3.8 & 19.75$\pm$0.05\\
 23.063 & Asiago & $R$ & 600 & 3.8 & 19.79$\pm$0.04\\
 23.069 & Asiago & $I$ & 600 & 3.8 & 19.27$\pm$0.06\\
 23.082 & AZT-24 & $K$ &2400 & 4.0 & 17.2$\pm$0.3$^2$\\
 23.083 & Asiago & $I$ & 600 & 3.8 & 19.32$\pm$0.06\\
 23.173 & Loiano & $B$ &2400 & 4.5 & 20.72$\pm$0.08$^3$\\
 23.199 & Loiano & $V$ &1800 & 4.0 & 20.37$\pm$0.10$^3$\\
 23.204 & AZT-24 & $J$ & 900 & 4.0 & 18.4$\pm$0.3$^2$\\
 23.211 & TNG    & $R$ & 120 & 1.1 & 20.00$\pm$0.01\\
 23.219 & TNG    & $R$ & 120 & 1.1 & 20.01$\pm$0.01\\  
 23.280 & TNG    & $U$ & 300 & 1.1 & 20.34$\pm$0.03\\
 23.284 & TNG    & $U$ & 300 & 1.1 & 20.33$\pm$0.03\\
 24.097 & TIRGO  & $J$ &4680 & 3.8 & 19.21$\pm$0.35\\
 24.106 & Loiano & $V$ &2400 & 2.5 & 21.23$\pm$0.06\\
 24.117 & TIRGO  &$K_s$&4680 & 3.3 & 17.49$\pm$0.25\\
 24.127 & Loiano & $R$ & 900 & 2.0 & 20.89$\pm$0.09\\
 24.236 & TNG    & $R$ & 120 & 1.0 & 21.06$\pm$0.03\\
 24.239 & TNG    & $R$ & 120 & 1.0 & 21.05$\pm$0.02\\
 24.241 & TNG    & $I$ & 120 & 1.0 & 20.42$\pm$0.04\\
 24.244 & TNG    & $I$ & 120 & 1.0 & 20.51$\pm$0.04\\
 24.247 & TNG    & $V$ & 120 & 1.0 & 21.44$\pm$0.02\\
 24.249 & TNG    & $V$ & 120 & 1.0 & 21.48$\pm$0.03\\
 24.252 & TNG    & $B$ & 300 & 1.0 & 21.87$\pm$0.02\\
 24.257 & TNG    & $B$ & 300 & 1.0 & 21.88$\pm$0.02\\
 24.262 & TNG    & $U$ & 450 & 1.0 & 21.33$\pm$0.04\\
 24.268 & TNG    & $U$ & 450 & 1.0 & 21.30$\pm$0.04\\
 25.253 & TNG    & $R$ & 120 & 0.9 & 21.64$\pm$0.03\\
 25.261 & TNG    & $V$ & 120 & 0.9 & 21.92$\pm$0.04\\
 25.263 & TNG    & $V$ & 120 & 0.9 & 21.96$\pm$0.03\\
 25.267 & TNG    & $B$ & 300 & 0.9 & 22.43$\pm$0.02\\
 25.271 & TNG    & $B$ & 300 & 0.9 & 22.47$\pm$0.03\\
 25.275 & TNG    & $I$ & 120 & 0.9 & 21.14$\pm$0.06\\
 25.277 & TNG    & $I$ & 120 & 0.9 & 21.16$\pm$0.06\\
 27.139 & Asiago & $R$ & 900 & 3.1 & 22.11$\pm$0.25\\
 27.264 & TNG    & $R$ & 360 & 1.2 & 22.38$\pm$0.05\\
 27.270 & TNG    & $V$ & 360 & 1.1 & 22.80$\pm$0.06\\
 27.278 & TNG    & $B$ & 360 & 1.1 & 23.28$\pm$0.08\\
Mar 31.226 & TNG & $R$ &3300 & 1.2 & 25.1$\pm$0.2\\
\noalign{\smallskip}
\hline
\noalign{\smallskip}
\multicolumn{6}{l}{$^1$Magnitudes of the GRB counterpart, not corrected
for interstellar absorption}\\
\multicolumn{6}{l}{$^2$This measurement supersedes the value
reported by Di Paola et al. (2001)}\\
\multicolumn{6}{l}{$^3$This measurement supersedes the value
reported by Bartolini et al. (2001)}\\
\end{tabular}
\end{center}
\end{table*}

\begin{table*}
\caption[]{List of the absorption lines identified in the TNG optical
spectrum of GRB010222 afterglow. The number associated with each line
refers to the identification shown in Fig. 4.
The error on all line positions can conservatively be assumed to be $\pm$3
\AA, i.e. comparable with the pixel size (see text). EWs of lines in the
GRB rest frame, i.e. divided by a factor (1+$z$), are also reported}

\begin{center}
\begin{tabular}{rcclcl}
\noalign{\smallskip}
\hline
\noalign{\smallskip}
\multicolumn{1}{c}{Line} & Observed & Rest frame &
\multicolumn{1}{c}{Element} & Redshift & \multicolumn{1}{c}{W$_{\rm
rest}$}\\
\multicolumn{1}{c}{number} & wavelength (\AA) & wavelength (\AA) & & &
\multicolumn{1}{c}{(\AA)}\\
\noalign{\smallskip}
\hline
\noalign{\smallskip}
1& 3588 & 1862.790  & Al{\sc iii}  &   0.926$\pm$0.002 & 2.1$\pm$0.2\\
2& 3979 & 2062.664  & Zn{\sc ii}/Cr{\sc ii} blend &
  0.929$\pm$0.001&2.5$\pm$0.3\\
3& 4573 & 2374.461  & Fe{\sc ii}   &   0.926$\pm$0.001 & 0.4$\pm$0.2\\
4& 4590 & 2382.765  & Fe{\sc ii}   &   0.926$\pm$0.001 & 1.2$\pm$0.3\\
5& 5014 & 2600.173  & Fe{\sc ii}   &   0.928$\pm$0.001 & 0.7$\pm$0.3$^*$\\
6& 5384 & 2796.352  & Mg{\sc ii}   &   0.925$\pm$0.001 & 0.7$\pm$0.3\\
7& 5401 & 2803.531  & Mg{\sc ii}   &   0.926$\pm$0.001 & 1.1$\pm$0.3\\
8& 5500 & 2852.964  & Mg{\sc i}    &   0.928$\pm$0.001 & 0.6$\pm$0.3\\
\noalign{\smallskip}
\multicolumn{3}{r}{Weighted mean} & ........................... &
0.927$\pm$0.001 &\\
\noalign{\smallskip}
\hline
\noalign{\smallskip}
9&  3599 & 1670.787 &   Al{\sc ii}  &    1.154$\pm$0.002 & 1.3$\pm$0.4\\
10& 3995 & 1854.716 &   Al{\sc iii} &    1.154$\pm$0.002 & 0.8$\pm$0.3\\
11& 5060 & 2344.214 &   Fe{\sc ii}  &    1.158$\pm$0.001 & 0.7$\pm$0.3\\
12& 5133 & 2382.765 &   Fe{\sc ii}  &    1.154$\pm$0.001 & 0.6$\pm$0.3\\
13& 5573 & 2586.650 &   Fe{\sc ii}  &    1.154$\pm$0.001 & 1.1$\pm$0.4\\
14& 5603 & 2600.173 &   Fe{\sc ii}  &    1.155$\pm$0.001 & 1.0$\pm$0.4\\
15& 6024 & 2796.352 &   Mg{\sc ii}  &    1.154$\pm$0.001 & 0.7$\pm$0.3\\
16& 6038 & 2803.531 &   Mg{\sc ii}  &    1.154$\pm$0.001 & 1.2$\pm$0.4\\
\noalign{\smallskip}
\multicolumn{3}{r}{Weighted mean} & ........................... &
1.155$\pm$0.001&\\
\noalign{\smallskip}
\hline
\noalign{\smallskip}
17&  3783 & 1526.707 &   Si{\sc ii} &     1.478$\pm$0.002 & 1.2$\pm$0.3\\
18&  3838 & 1550.774 &   C{\sc iv}  &     1.475$\pm$0.002 & 2.3$\pm$0.2\\
19&  4132 & 1670.787 &   Al{\sc ii} &     1.473$\pm$0.002 & 1.1$\pm$0.3\\
20&  4472 & 1808.013 &   Si{\sc ii} &     1.474$\pm$0.002 & 0.7$\pm$0.3\\
21& 5014 & 2026.136 & Zn{\sc ii}/Cr{\sc ii} blend & 1.475$\pm$0.001&
  0.6$\pm$0.2$^*$\\
22& 5105 & 2062.664 & Zn{\sc ii}/Mg{\sc i} blend  & 1.475$\pm$0.001&
  1.1$\pm$0.3\\
23&  5800 & 2344.214 &   Fe{\sc ii} &     1.474$\pm$0.001 & 2.2$\pm$0.2\\
24&  5880 & 2374.461 &   Fe{\sc ii} &     1.476$\pm$0.001 & 1.9$\pm$0.2\\
25&  5894 & 2382.765 &   Fe{\sc ii} &     1.474$\pm$0.001 & 1.7$\pm$0.2\\
26&  6384 & 2576.107 &   Mn{\sc ii} &     1.478$\pm$0.001 & 0.5$\pm$0.2\\
27&  6401 & 2586.650 &   Fe{\sc ii} &     1.474$\pm$0.001 & 0.8$\pm$0.3\\
28&  6423 & 2593.731 &   Mn{\sc ii} &     1.476$\pm$0.001 & 2.0$\pm$0.2\\
29&  6435 & 2600.173 &   Fe{\sc ii} &     1.475$\pm$0.001 & 1.9$\pm$0.2\\
30&  6454 & 2605.697 &   Mn{\sc ii} &     1.477$\pm$0.001 & 1.1$\pm$0.3\\
31&  6919 & 2796.352 &   Mg{\sc ii} &     1.474$\pm$0.001 & 2.8$\pm$0.2\\
32&  6937 & 2803.531 &   Mg{\sc ii} &     1.474$\pm$0.001 & 3.4$\pm$0.3\\
33&  7069 & 2852.964 &   Mg{\sc i}  &     1.478$\pm$0.001 & 1.0$\pm$0.3\\
\noalign{\smallskip}
\multicolumn{3}{r}{Weighted mean} & ........................... &
1.475$\pm$0.001&\\
\noalign{\smallskip}
\hline
\noalign{\smallskip}
\multicolumn{5}{l}{$^*$These lines might be blended with each other}\\
\end{tabular}
\end{center}
\end{table*}

\section{Observations and data reduction}

We observed the GRB010222 field at the 3.58-meter {\it Telescopio
Nazionale Galileo} (TNG) in the Canary Islands (Spain), at the 1.8-meter
``Copernicus" telescope at Cima Ekar of the ``Leonida Rosino" Astronomical
Observatory of Asiago (Italy), and at the University of Bologna 1.52-meter
``G.D. Cassini" telescope in Loiano (Italy). 

TNG was equipped with the spectrophotometer DOLoReS and a 2048$\times$2048
pixels Loral CCD which allows covering a 9$\farcm$5$\times$9$\farcm$5
field in imaging mode with a scale of 0$\farcs$275/pix; the ``Copernicus"
telescope was carrying the AFOSC instrument whose 1100$\times$1100 pixels
SITE CCD has a field of view of 8$\farcm$5$\times$8$\farcm$5 with a scale
of 0$\farcs$47/pix; the ``Cassini" telescope was mounting the 
spectrophotometer BFOSC with a 1340$\times$1340 pixels EEV CCD, a field of
view of 12$\farcm$5$\times$12$\farcm$5 and a scale of 0$\farcs$58/pix.

NIR imaging in $J$ and $K$ bands was acquired at the AZT-24 1.1-meter
telescope at Campo Imperatore (Italy) with SWIRCAM, a 256$\times$256
pixels infrared camera with a field of view of
4$\farcm$4$\times$4$\farcm$4 and
a pixel scale of 1$\farcs$03/pix, and in $J$ and $K_s$ at the 1.5-meter
Gornergrat infrared telescope TIRGO (Switzerland) using the 256$\times$256
pixels infrared camera ARNICA which has a 4$'$$\times$4$'$ field of view
and a pixel scale of 0$\farcs$97/pix. The $K_s$ filter is centered at 2.12
$\mu$m and has a full width at half maximum of 0.34 $\mu$m.
For each NIR observation the total integration time was split into
images of 30 s each, and the telescope was dithered in between.

The complete log of our imaging observations is reported in Table 1.

Two 30-min spectra, nominally covering the 3000-8000 \AA~optical range,
were also acquired with TNG+DOLoReS between Feb 23.227 and Feb
23.274 UT. The use of DOLoReS Grism \#1 secured a dispersion of 2.4
\AA/pix. The slit width was 1$\farcs$5.

\begin{figure*}
\begin{center}
\epsfig{figure=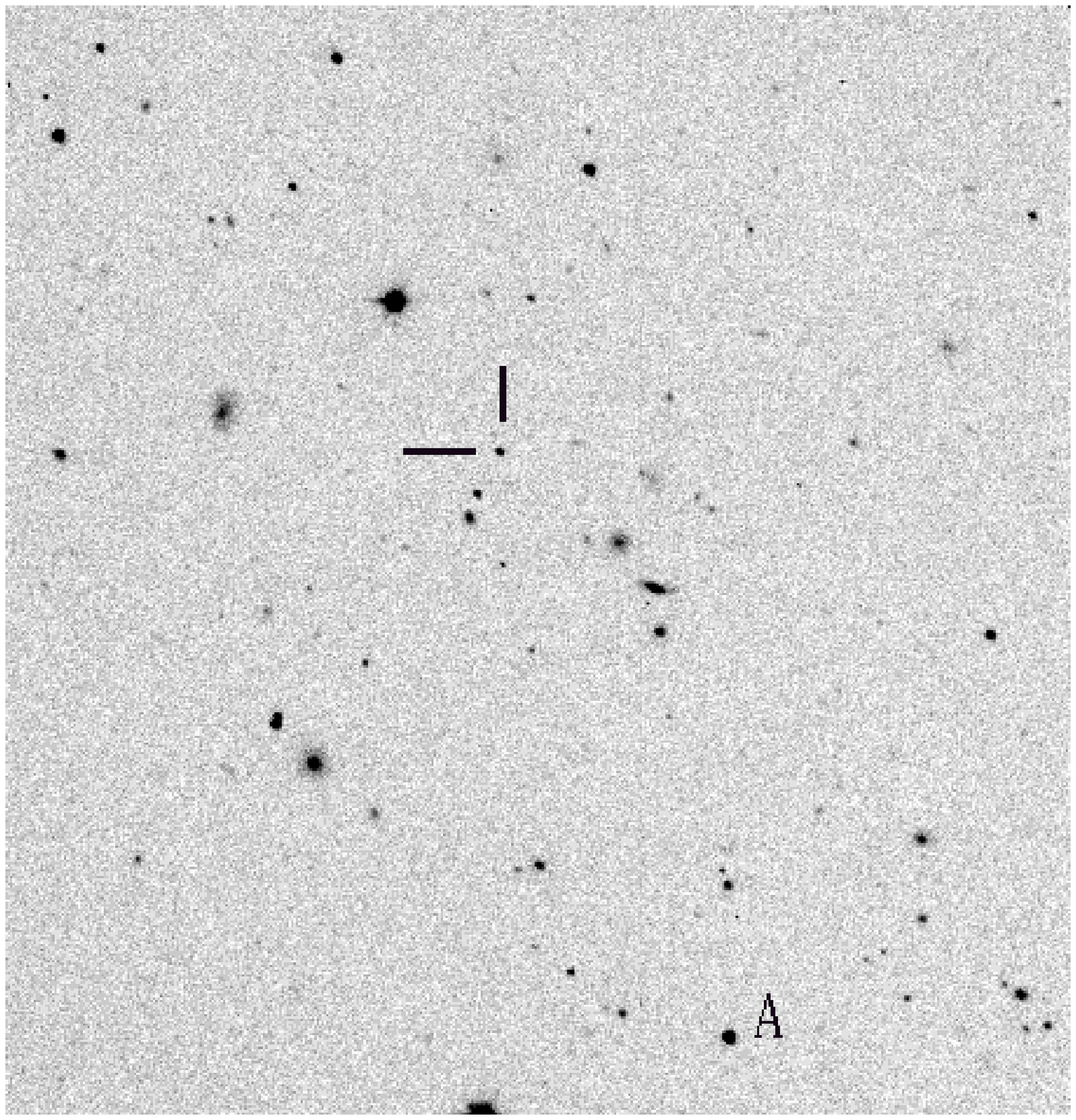,width=8.4cm}
\epsfig{figure=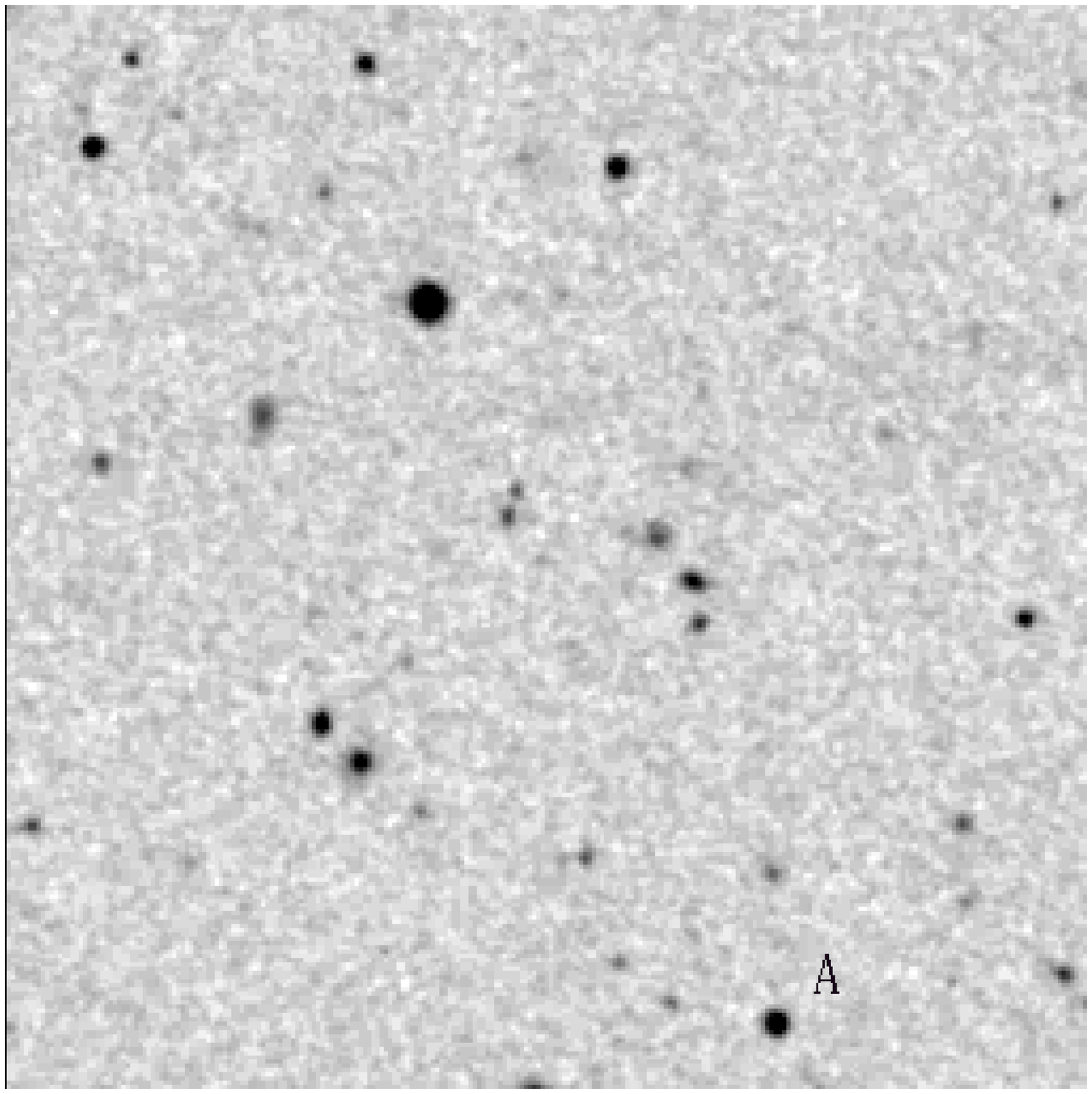,width=8.7cm}
\end{center}
\vspace{-.5cm}
\caption[]{{\it (Left panel)} Field of GRB010222 as imaged by the TNG on
Feb 23.219 UT in the $R$ band. The OT is indicated by the tickmarks.
{\it (Right panel)} The same field as it appears on the DSS-II sky survey.
No source is detected at the OT position.
Both fields have a size of about 4$'$$\times$4$'$; North is at top and
East is to the left. We also indicated the star `A' by Stanek et al.
(2001a), which we used for the calibration}
\end{figure*}

\begin{figure}
\begin{center}
\epsfig{figure=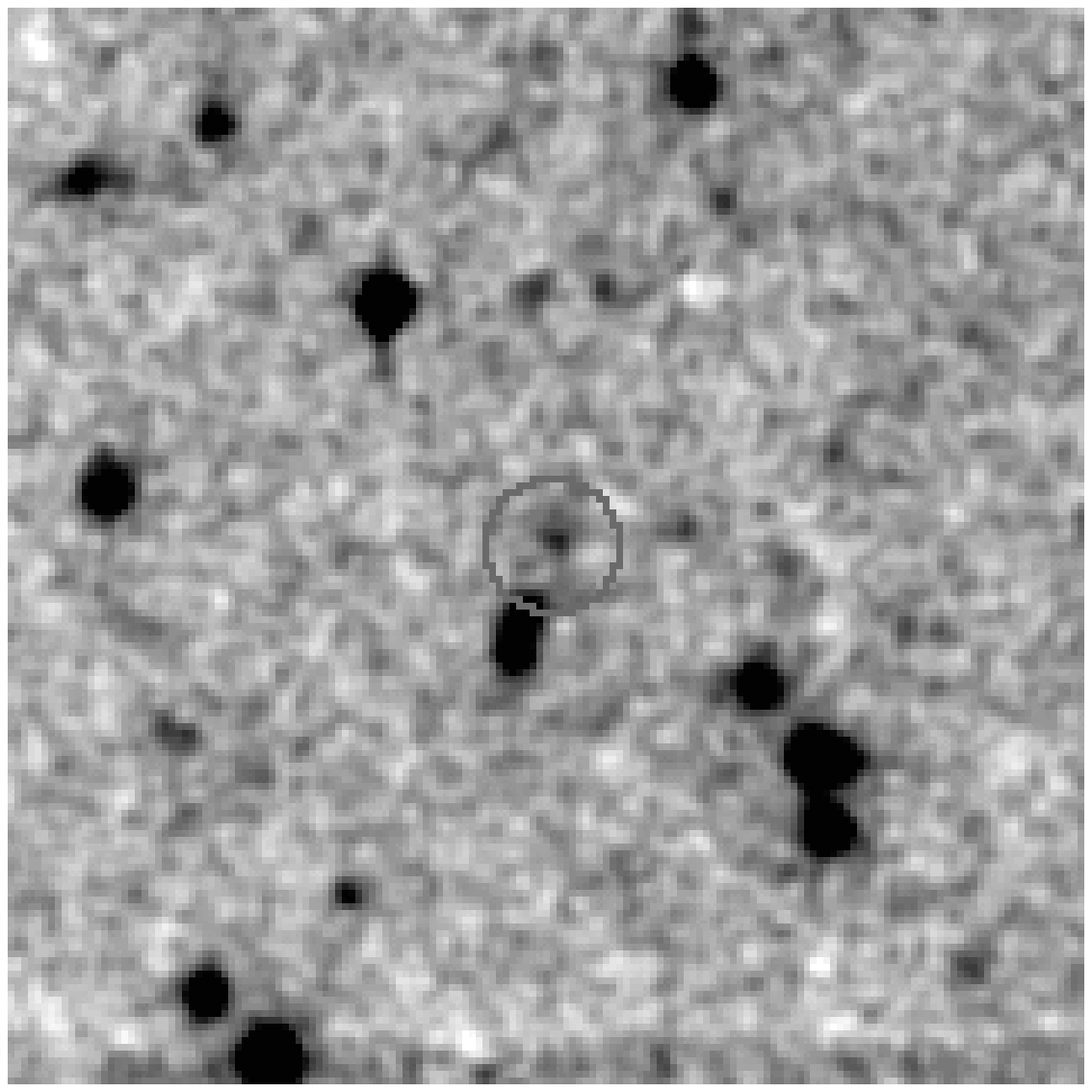,width=9cm}
\end{center}
\vspace{-.8cm}
\caption[]{Smoothed $K_s$-band image acquired at TIRGO on 2001 Feb
24.117 UT and centered at the NIR afterglow position. The image size is
about 2$\farcm$5$\times$2$\farcm$5. North is at top, east is to the left.
The circle indicating the position of the NIR transient has a radius of
15$''$}
\end{figure}

\medskip

Optical images were debiased and flat-fielded with the standard cleaning
procedure. In some cases, frames taken on the same night in the same band
were summed together in order to increase the signal-to-noise ratio.
In Fig. 1 we report our second TNG $R$-band image.
We chose standard Point Spread Function (PSF) fitting as photometric
method, and to this aim we used the {\sl DAOPHOT II} image data
analysis package PSF-fitting algorithm (Stetson 1987) running within 
MIDAS\footnote{MIDAS (Munich Image Data Analysis System) is
developed, distributed and maintained by ESO (European Southern
Observatory) and is available at {\tt http://www.eso.org/projects/esomidas}}.
A two-dimensional Gaussian profile with two free parameters (the half
width at half maxima along $x$ and $y$ coordinates of each frame) was
modeled on at least 5 unsaturated bright stars in each image. 
The errors associated with the measurements reported in Table 1 represent 
statistical uncertainties (at 1$\sigma$) obtained with the
standard PSF-fitting procedure.
In only one case, i.e. the TNG observation of Mar 31, we used
aperture photometry for the magnitude determination as the PSF-fitting
procedure could not give reliable results due to the faintness of the
transient. In this case we used an aperture diameter equal to the seeing
FWHM of the summed image.

To be consistent with magnitude measurements appeared on the GCN
circulars archive\footnote{GCN Circulars are
available at:\\ {\tt http://gcn.gsfc.nasa.gov/gcn/gcn3\_archive.html}},
calibration was done using the $UBVRI$ magnitudes, as measured by Henden
(2001c), of field star `A' indicated by Stanek et al. (2001a). However we
used other field stars to check the stability of this calibration: we
found it to be accurate to within 5\%. Any significant short-term
variation of star `A' can be ruled out from our data.
Due to the $U-B$ color dependence of the DOLoReS CCD response in the
$U$ band, a $\sim$0.2 mag color-term correction was subtracted from the
$U$ magnitudes of the OT.
We remark that the photometry errors quoted throughout the rest of the
paper are only statistical and do not account for any possible 
(most likely very small) zero point offset.

We also retrieved and reduced the two $V$-band images obtained by Billings
(2001) on 2001 Feb. 22.502 and 22.547 UT. From these, we obtain $V$ =
18.80$\pm$0.09 and 19.11$\pm$0.11, respectively, assuming the calibration 
by Henden (2001c) as described above.

Standard procedures were used to reduce the NIR data:
a sky estimate for each image was computed from the clipped median
of the nearby images. The telescope dithering was measured from the
offsets of field objects in each image and the images were averaged
together using inter-pixel shifts.
Magnitudes were measured inside circular apertures of diameter 10$''$
and corrected to total magnitudes.
The photometric calibration was performed using NIR standard stars from
the list by Hunt et al. (1998). 

Given the poor seeing, the objects within
10$''$ from the OT location could contribute to the measured NIR flux of
the transient. It is not possible to accurately measure this effect, but
we estimate that it is at most a small fraction of the total. Indeed, the
peak of the emission is in good agreement with the expected position of
the OT, and no evidence of flux is seen at the position of an object
detected in the optical at about 5$''$ north of the OT. 
Furthermore, the PSF profiles of the OT and of the object located 13$''$
south-east of the OT itself are well distincted, suggesting that the
latter does not contaminate the OT flux (see Fig. 2).

We rescaled the TIRGO $K_s$ magnitude to
the standard $K$ band. This was done in two steps: we first considered
that, given the almost identical characteristics of the $K_s$ and $K'$
filters (Wainscoat \& Cowie 1992) we could assume $K'$ = $K_s$ within the
large errors associated with our $K_s$ measurement. Second,
using the relation $K' - K$ = 0.2 $\times$ $(H-K)$ by Wainscoat \& Cowie
(1992) and assuming the power law spectral shape described in Section 3.3
to estimate the $H$-band magnitude of the OT, we obtained $K' - K$
$\simeq$ 0.15. Thus, $K$ = 17.35$\pm$0.3 at the time of the TIRGO
observation.

We then evaluated the Galactic absorption in the optical and NIR bands
along the direction of GRB010222 using the Galactic dust infrared maps
by Schlegel et al. (1998); from these data we obtained a color excess
$E(B-V)$ = 0.022. By applying the law by Rieke \& Lebofsky (1985), this color
excess corresponds to $A_V$ = 0.07; then, using the relation by Cardelli
et al. (1989), we derived $A_U$ = 0.12, $A_B$ = 0.09, $A_R$ = 0.05,
$A_I$ = 0.04, $A_J$ = 0.02, $A_K$ = 0.01.

Spectra, after correction for flat-field and bias, were background
subtracted and optimally extracted (Horne 1986)
using IRAF\footnote{IRAF is the Image Reduction and Analysis Facility made
available at {\tt http://iraf.noao.edu} to the astronomical
community by the National Optical Astronomy Observatories, which are
operated by AURA, Inc., under contract with the U.S. National Science
Foundation. STSDAS is distributed by the Space Telescope Science
Institute, which is operated by the Association of Universities for
Research in Astronomy (AURA), Inc., under NASA contract NAS 5--26555.}.
Helium-Argon lamps were used for wavelength calibration; spectra were then
flux-calibrated by using the spectrophotometric standard Feige 34 (Massey
et al. 1988) and finally averaged together. 
The correctness of the wavelength and flux calibrations was checked
against the position of night sky lines and the photometric data collected
around the epoch in which the spectra were acquired, respectively.
The typical error was 0.5 \AA~for the wavelength calibration and 5\% for
the flux calibration.

\begin{figure*}
\vspace{-1.5cm}
\begin{center}
\epsfig{figure=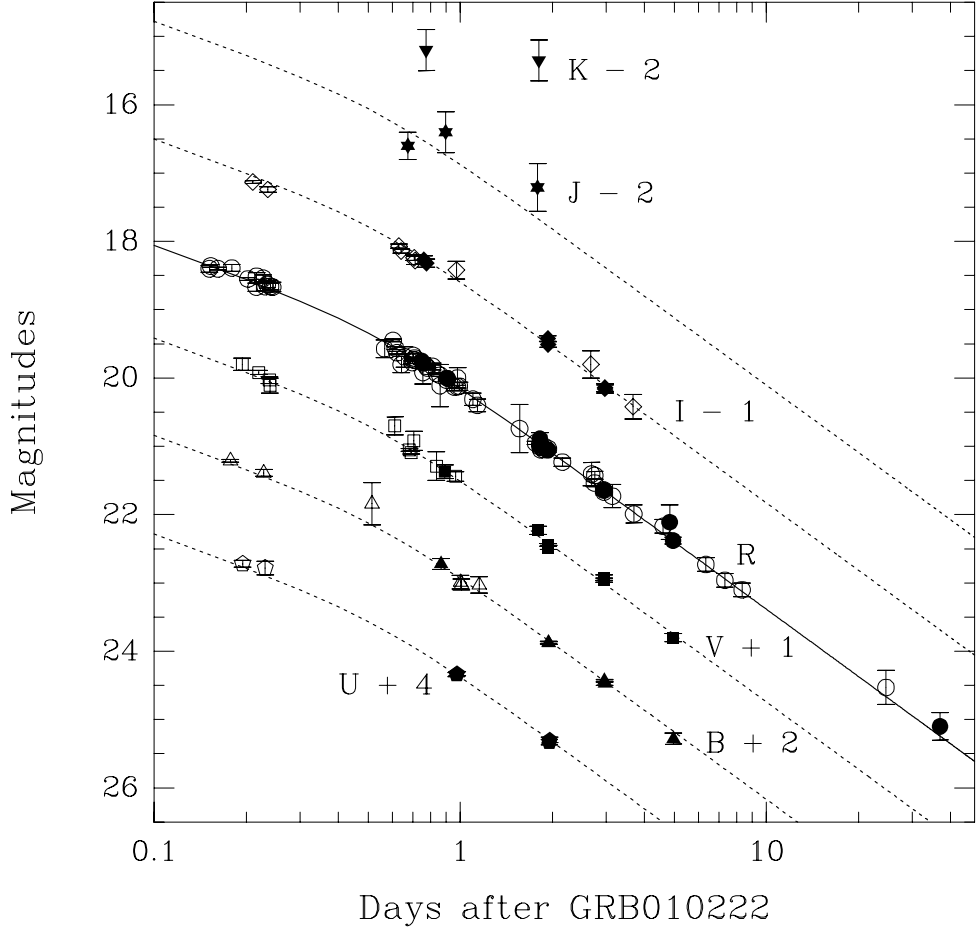,width=16cm}
\end{center}
\vspace{-1.8cm}
\caption[]{$UBVRIJK$ light curves of GRB010222 afterglow, based on the
data presented in this paper and in the literature (see text). 
Different symbol styles indicate different bands. Filled symbols represent
data presented in this work, while open symbols refer to measurements
published by other authors. 
We have consistently referred all optical magnitudes to the calibration
zero point of Henden (2001c).
No Galactic extinction correction, nor host galaxy flux subtraction have
been applied. The GRB start time corresponds to 2001 February 22.30799
UT. Overplotted to the $R$-band data (solid curve) is the best-fit double
power law empirical model described in Sect. 3.1. Fit parameters are:
$\alpha_1$ = 0.65$\pm$0.15, $\alpha_2$ = 1.31$\pm$0.02, 
$t_b$ = 0.55$\pm$0.15 days.
The same model curve, with unchanged model parameters, except for 
the flux normalization at the break time, accounts well for
the data in the other bands (dashed curves), except $K$}
\end{figure*}

\begin{figure*}
\vspace{-7cm}
\begin{center}
\hspace{1cm}
\epsfig{figure=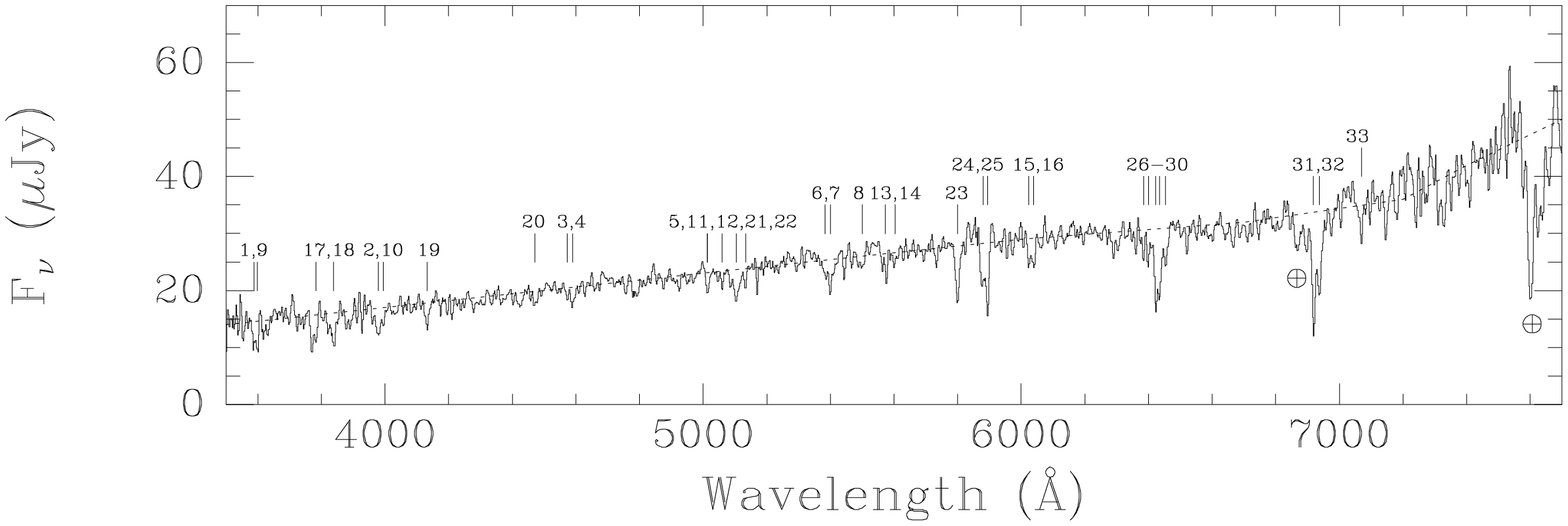,angle=-90,width=16cm}
\end{center}
\vspace{-7.5cm}
\caption[]{TNG+DOLoReS spectrum of the GRB010222 afterglow in the
3500-7700 \AA~range, smoothed with a Gaussian filter with
$\sigma$=3 \AA~(i.e. comparable with the spectral dispersion) and
corrected for Galactic absorption assuming $E(B-V)$ = 0.022.
Numbers mark the positions of identified lines as listed in Table 2. 
Telluric absorption lines at 6870 and 7600 \AA~are also marked with the
symbol $\oplus$. The dashed curve indicates the position of the
spectral continuum we used to measure the EW of lines}
\end{figure*}

\section{Results}

\subsection{Photometry}

We clearly detect the transient in all optical and NIR filters; Fig. 1
shows the TNG $R$-band image taken on 2001 Feb. 23.219 UT along with the
same portion of sky as it appears on the DSS-II survey. The OT,
indicated by the tickmarks, is clearly evident on the TNG image, while
no object is present in the DSS-II image at that position. The association
of this transient with GRB010222 is confirmed by the fading behaviour of
this source (see optical magnitudes reported in Table 1).
The light curves in $UBVRI$ bands are reported in Fig. 3, where our data
are complemented with those published by other authors (GCN circulars
archive; Sagar et al. 2001; Stanek et al. 2001c; Cowsik \& Bhargavi 2001).
Note that we rescaled the $BVRI$ magnitudes appeared in the GCN circulars
to the zero point measured by Henden (2001c); we did not plot in Fig. 3
the measurements in GCN circulars which could not be tied to this
photometric zero point (for instance, unfiltered or independently 
calibrated magnitudes).
No correction has been applied for Galactic extinction, which is anyway
small (see previous Section), to the data in Table 1 and Fig. 3; nor 
has any host galaxy continuum emission been subtracted, this being
unknown at present.

We find that a single power law decay of the form $F \propto
(t-t_0)^{-\alpha}$, where $t_0$ is the GRB start time, does not provide a
satisfactory fit of the light curves. Indeed, considering a decay index
$\alpha\sim$ 0.9 as reported by Price et al. (2001b) and Fynbo et al. 
(2001b), we obtain a fit with $\chi^2_\nu\sim$ 65 over 68 degrees of
freedom (dof). With this fit, the measurements acquired starting about 1
day after the GRB fall systematically below the expected decay. This is
clearly evident in the $B$-, $V$- and $R$-band light curves, the best
sampled since the first hours after the GRB.

We thus hypothesize that the $R$-band light curve can be modeled with a 
smoothly broken power law as the one used by Stanek et al. (1999)
to fit the optical light curves of the GRB990510 afterglow. The model we
apply is:
\begin{equation}
F_\nu (t) =
\frac{k_\nu}{(\frac{t}{t_b})^{-\alpha_1}+(\frac{t}{t_b})^{-\alpha_2}}~,
\end{equation}
where $t_b$ is the time of the light curve break, $\alpha_1$ is the
asymptotic decay index for $t\ll t_b$ and $\alpha_2$ is the
asymptotic decay index for $t\gg t_b$ (times are expressed since $t_0$).
Note that for the temporal decay indices we use inverted notations with
respect to those of Stanek et al. (1999). We obtain a best fit 
of the $R$-band data with $\alpha_1$ = 0.23$\pm$0.01, $\alpha_2$ =
1.57$\pm$0.01 and $t_b$ = 0.54$\pm$0.01 days after the GRB start time;
however, this fit is not satisfactory, as it has $\chi^2_\nu\sim$ 14
(66 dof) and systematically underestimates the OT flux for $t >$ 5 days
after the GRB.

We next try the generalization adopted by Beuermann et al. (1999)
of a smoothly broken power law model:
\begin{equation}
F_\nu (t) = (F_1(t)^{-s}+F_2(t)^{-s})^{-\frac{1}{s}}~,
\end{equation}
where $F_i(t) = k_i t^{-\alpha_i}$ (with $i$ = 1,2), and in which
the parameter $s$ indicates the smoothness of the change from a decay
index to the other; for very large values of $s$, this function assumes a
broken power law shape.
This model fits much better ($\chi^2_\nu$ = 1.6, 65 dof) the $R$ points
than model of Eq. (1). 
An F-test shows that the model by Beuermann et al. (1999) leads to a tiny
chance probability of improvement with respect to model of Eq. (1).

The formal fit parameters are: $\alpha_1$ = 0.54$\pm$0.08, $\alpha_2$ =
1.31$\pm$0.02, $t_b$ = 0.44$\pm$0.05 days after the GRB start time, and
`smoothness parameter' $s$ = 3.1$\pm$1.0. The value of this
parameter indicates that the slope change occurred quite fastly around 12
hours after $t_0$. 
However, the fit parameters $\alpha_1$ and $t_b$ are very sensitive
to the value of $s$, and we have noted that similarly acceptable fits 
($\chi^2_\nu$ ranging from 1.6 to 1.7) are obtained when $s$ varies 
from $\sim$3 to $\sim$10.
Correspondingly, $\alpha_1$ varies between $\sim$0.5 and $\sim$0.8, and
$t_b$ between $\sim$0.4 and $\sim$0.7 days.
The upper bounds of the intervals for these parameters are more
consistent with the results of other authors (Sagar et al. 2001;
Stanek et al. 2001c). 
Therefore, we will assume for $\alpha_1$ and $t_b$ their average values
within the above ranges, and the associated dispersions as uncertainties:
$\alpha_1$ = 0.65$\pm$0.15 and $t_b$ = 0.55$\pm$0.15 days.
The parameter $\alpha_2$ is instead very weakly sensitive to the value of
$s$. 

The $\chi^2_\nu$ values associated with the fits are larger than 1 and
therefore formally not completely satisfactory.
By applying a systematic error of 1.5 \% to all $R$ data points, in
addition to the statistical and calibration uncertainties, the
$\chi^2_\nu$ becomes $\sim$1. Such a small additional uncertainty can be
ascribed to non-homogeneity of $R$-band data set (acquired at different
telescopes and analyzed using different methods), although small
irregularities in the circumburst medium may produce in the afterglow
light curve small time scale fluctuations of similar amplitude
(e.g. GRB000301C, Masetti et al. 2000b; see also Wang \& Loeb 2000).

The $U$ $B$, $V$, $I$ and $J$ light curves, albeit less well sampled, are
consistent with the model describing the $R$-band data (see Fig. 3);
therefore, no appreciable color evolution is present in the optical
afterglow. On the contrary, the $K$-band light curve is consistent with
being constant between the two epochs of observation.

From the fits, we obtained the following average color
indices (not corrected for Galactic absorption): $U-B$ = $-$0.55$\pm$0.05,
$B-V$ = 0.43$\pm$0.05, $V-R$ = 0.36$\pm$0.05, $R-I$ = 0.55$\pm$0.05
and $I-J$ = 0.72$\pm$0.15.
These figures place the OT of GRB010222 in the locus populated by GRB
optical afterglows in the optical color-color diagrams, as illustrated by
\v{S}imon et al. (2001). 

\subsection{Spectroscopy}

We detect several absorption features in the optical spectrum of the
GRB010222 afterglow (Fig. 4). As earlier reported by other authors
(Garnavich et al. 2001; Jha et al. 2001a,b; Bloom et al. 2001), these
correspond to three different metal absorption systems located at
different redshifts. 
Our line fitting, performed with the SPLOT task within IRAF, assumes a
Gaussian profile for the absorption lines. A conservative error of $\pm$3
\AA, comparable with the pixel size, is associated with each line
wavelength measurement.

On the averaged TNG spectrum, presented in Fig. 4, we mark the positions
of all lines we identified. Each number corresponds to a line
in Table 2, where line wavelengths, identifications and redshifts are
reported. From our line identifications we measure
the three absorption systems at $z$ = 0.927$\pm$0.001, $z$ =
1.155$\pm$0.001 and $z$ = 1.475$\pm$0.001, the highest one thus being the
lower limit for the redshift of this GRB. These results are consistent
with those of Garnavich et al. (2001), Jha et al. (2001a,b) and Bloom et
al. (2001). Our spectral resolution is not sufficient to confirm the
presence of a fine structure in the farthest absorbing system as reported
by Castro et al. (2001).

In Table 2 we also list the equivalent widths (EWs) of the identified 
lines computed in the absorber rest frame (i.e. by dividing the measured
value by the factor 1+$z$). The errors on the EWs are computed by assuming
different spectral continuum levels roughly corresponding to 1$\sigma$
variation of the continuum itself in proximity of each line.

It can be noted from Table 2 that the metal lines of the absorption system
at the highest $z$ are very strong. In particular, the detection of 
Mg{\sc i} suggests that the optical emission of the GRB afterglow pierced
through a dense medium, most likely that of a host galaxy at $z$=1.475.
If we consider the study of $z <$ 1.65 Mg{\sc ii} systems
in QSO spectra by Rao \& Turnshek (2000) we find that, according to these
authors, Mg{\sc ii} systems with rest frame EW of Mg{\sc ii}$\lambda$2796
$\geq 0.6$ \AA~are primarily associated with
Damped Lyman--$\alpha$ Absorption (DLA) systems. DLAs are characterized by
being mostly H{\sc i} interstellar medium clouds of high redshift
galaxies.
Indeed, about 50\% of the Rao \& Turnshek's (2000) sample of Mg{\sc ii}
absorbers with Mg{\sc ii}$\lambda$2796 and Fe{\sc ii}$\lambda$2600
EWs larger than 0.5 \AA~have H{\sc i} column density 
$N_{\rm HI}>2\times 10^{20}$ atoms cm$^{-2}$ and basically all have
$N_{\rm HI}>10^{19}$ atoms cm$^{-2}$.  

Moreover, they found that almost all DLAs with $N_{\rm HI}>2 \times
10^{20}$ atoms cm$^{-2}$ have Mg{\sc i}$\lambda$2852 EW larger than
0.7 \AA. Thus, given the EWs of Mg{\sc i}, Fe{\sc ii} and 
Mg{\sc ii} reported in Table 2, we have good indications that the
$z$=1.475 absorber, if not associated with the GRB itself, most
likely resembles the properties of a DLA system with column density
$N_{\rm HI}>2 \times 10^{20}$ atoms cm$^{-2}$. The ratio of the
EWs of the Mg{\sc ii} doublet is 
$W_{\rm rest}(\lambda2796)/W_{\rm rest}(\lambda2803)$ = 0.8$\pm$0.1; this
indicates that the absorption is saturated and, given the low resolution
of the spectrum, the column density is undetermined. Rao \& Turnshek
(2000) also show that there is no precise relationship between the 
H{\sc i} column density and the rest frame EW of Mg{\sc ii}$\lambda$2796.
Thus, although we have good indications that the $z$=1.475 system bears
substantial absorption, the H{\sc i} column density is basically unknown.
Applying the same approach to the other two absorption systems lying
along the GRB line of sight (although in the case of the absorber at
$z$=1.155 no Mg{\sc i} is detected), we suggest that these systems have a
H{\sc i} column density in excess of $10^{19}$ atoms cm$^{-2}$.
All this points to the possibility of the presence of additional
extragalactic absorption along the GRB010222 line of sight.

Finally, we tested the hypothesis (Sari et al. 1998) that the optical
spectrum can be described with a power law in the form F$_\nu \propto
\nu^{-\beta}$. If we fit the TNG spectrum with this law, we
obtain $\beta$ = 1.16$\pm$0.05 (with $\chi^2_\nu \sim$ 1), 
consistent with the fit results from the multiwavelength spectra
shown in the next Subsection, but somewhat steeper than the value
reported by Jha et al. (2001b).

\subsection{Multiwavelength spectra}

In order to study the temporal evolution of the multiwavelength
spectrum of the GRB010222 afterglow, we have selected 5 epochs from
$\sim$0.2 to $\sim$5 days after the high-energy event, at which best
spectral sampling is available. We corrected the afterglow optical/NIR
magnitudes for the Galactic absorption and converted them into fluxes
using the tables by Fukugita et al. (1995) for the optical and by
Bersanelli et al. (1991) for the NIR.
If strictly simultaneous observations are not available, we
reduced the fluxes to the reference epoch by interpolating the empirical
model fitted to the $R$-band light curve (see Sect. 3.1), and normalizing
to the appropriate band.
A 7\% systematic error is added quadratically to the measurements
to account for interpolation and magnitude-to-flux conversion
uncertainties. The results are plotted in Fig. 5.

The optical spectral distributions covering $B$, $V$, $R$ and, when
available, $U$ and $I$ filters are well fit at all epochs by a single
power law of average spectral index $\beta$ = 1.1$\pm$0.1 (the
$\chi^2_\nu$ is typically $\sim$1). A constant spectral shape is
consistent with the achromatic evolution of the OT decay.  

A single power law fit to the $UBVRIJK$ data at the third epoch
gives $\beta$ = 1.3$\pm$0.1 with $\chi^2_\nu$ = 2.3
(5 dof), slightly steeper than the optical slope. Although this may
suggest some absorption within the host galaxy in addition to the Galactic
one (e.g., Lee et al. 2001), we note that the difference is barely
significant, considering the uncertainty on the joint optical/NIR spectral
fit. Therefore we conclude that the optical spectra at all epochs and the
optical/NIR spectrum at 1.8 days after the GRB do not deviate 
significantly from a single power law of slope $\beta \simeq 1$.

On the contrary, the $K$-band point at the second epoch ($\Delta t$ = 0.97
days after the GRB) strongly deviates from this trend. If we attempt to
fit all $UBVRIJK$ points at this epoch with a single power law we obtain
an unacceptable fit.

\begin{figure}
\begin{center}
\epsfig{figure=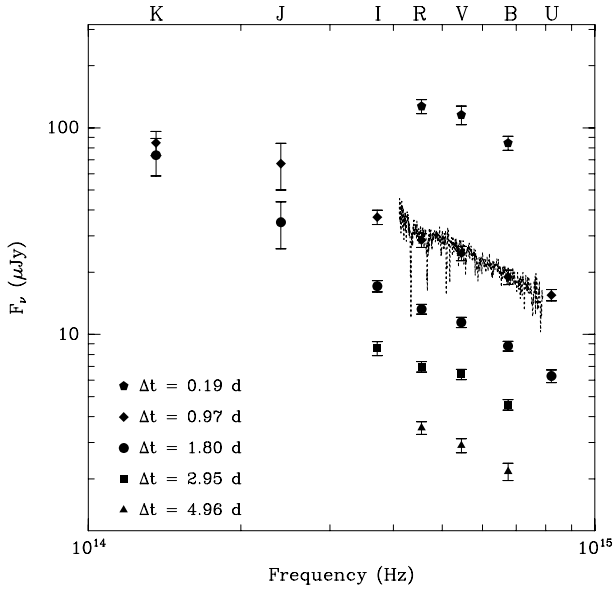,width=9.5cm}
\end{center}
\vspace{-1cm}
\caption[]{Optical/NIR broad-band spectra of the GRB010222
afterglow at five different epochs: 0.19, 0.97, 1.80, 2.95, and 4.96 days
after the GRB. The data are corrected for Galactic absorption. 
A 7\% systematic error was added quadratically to the
measurement errors in order to account for interpolation and 
magnitude-to-flux conversion uncertainties. 
For the sake of comparison, we also plotted the TNG optical spectrum
reported in Fig. 4 (acquired at epoch $\Delta t$ = 1.04 days after the
GRB)}
\end{figure}

\begin{figure}
\begin{center}
\epsfig{figure=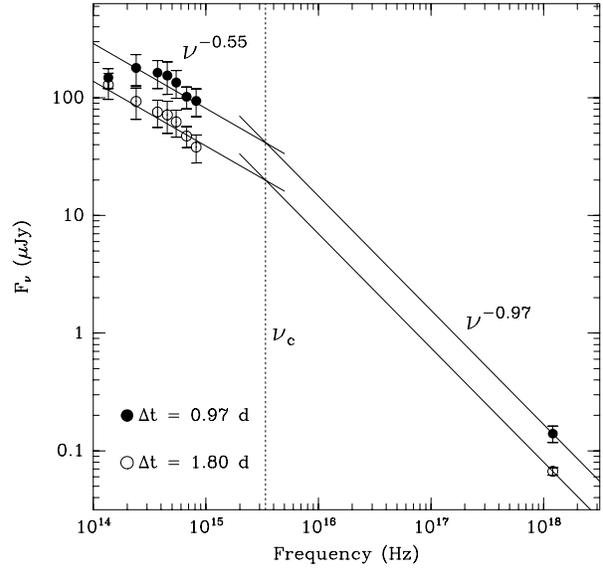,width=9.5cm}
\end{center}
\vspace{-1cm}
\caption[]{Multiwavelength X--ray/optical/NIR SEDs of the GRB010222
afterglow at epochs 0.97 and 1.8 days after the GRB. The X--ray data are
from in 't Zand et al. (2001). Optical and NIR data were corrected both
for Galactic extinction and for intrinsic host galaxy absorption using the
extinction curve by Calzetti (1997) and a local color excess $E(B-V)$ =
0.35. The resulting optical/NIR spectrum has a slope $\beta \sim$ 0.55 at
both epochs (the $K$-band point of the first epoch was excluded from the
fit because highly deviant). This plot indicates that $\nu_{\rm c}$ lies
between the optical and X--ray ranges, at $\simeq$3.5$\times$10$^{15}$ 
Hz}
\end{figure}

\section{Discussion}

The optical light curves of the GRB010222 afterglow are described
by a continuously steepening trend, instead of a single power law,
a behaviour seen in about 1/3 of the well monitored optical
afterglows (GRB980519, Jaunsen et al. 2001; GRB990123,
Castro-Tirado et al. 1999;  GRB990510, Stanek et al. 1999;
GRB990705, Masetti et al. 2000a; GRB991208, Castro-Tirado et al.
2001; GRB991216, Halpern et al. 2000; GRB000301C, Masetti et al.
2000b, Jensen et al. 2001; GRB000926, Fynbo et al. 2001a).  In the
present case, this cannot be ascribed to the synchrotron cooling
frequency $\nu_{\rm c}$ crossing the optical band in a spherical
regime (Sari et al. 1998), because the expected variation of decay
index would be in this case much smaller ($\Delta\alpha$ = 0.25)
than observed ($\Delta\alpha \simeq 0.75$).  Moreover, although the
limited optical band hampers a neat distinction, the $\nu_{\rm c}$
transition should be wavelength dependent, while the observed
steepening is achromatic.

Therefore we have considered a decelerating and sideways expanding
jet scenario (Sari et al. 1999). In this case, the index $\alpha_2$
should equal the index of the electron energy distribution, thus
$p$ = 1.3. From this follows that the optical decay index before
the collimation break time must be $\alpha_1$ = 0.23, in case
$\nu_{\rm c}$ is above the optical band at those epochs.  This is
rather different from our value of $\alpha_1$, 0.65$\pm$0.15.
Therefore, within the present picture, at the observing
epochs preceding the collimation break time, $\nu_{\rm c}$ should
have already crossed the optical band, so that the predicted
$\alpha_1$ would be $\sim$0.5, more similar to its observed value. 
The slope of the optical spectrum should then be $p$/2 $\simeq$
0.7, significantly flatter than both our measured value $\beta$ =
1.1 and that ($\beta$ = 0.9) reported by Jha et al. (2001b).

Local absorption, on top of the Galactic extinction, may
account for the difference (see below), but would lead to an
intrinsic optical spectral slope rather different than found in
X--rays ($\beta_{\rm X} \simeq 1$; in 't Zand et al. 2001), which is
inconsistent with the fact that both bands should be above $\nu_{\rm
c}$. In addition, the index of the electron energy distribution
derived in the sideways expanding jet scenario, $p$ = 1.3, is very
flat (in general, $p$ values larger than 2 are found; see Frontera
et al. 2000) and may pose energetic problems (although one can
assume that the electron power law has a cut-off at some energy).
Emission from the still undetected host galaxy of GRB010222 may
flatten the late epoch light curve of the afterglow, but the most
recent $R$-band photometry (our Fig. 3; Stanek et al. 2001c; Sagar
et al. 2001) suggests that the total magnitude of the host cannot be
brighter than $R \simeq 27$ and that its contribution at the latest
monitoring epochs ($\sim$40 days after the GRB) is not significant, 
so that subtracting it from the optical measurements does not alter the
fitted value of $\alpha_2$, and therefore of $p$.

Given the problems posed by the application of the above model to
our optical/NIR afterglow data, we have considered two alternatives. 

First we have compared our results with a model of an external shock
expanding in a pre-ejected wind, which determines a $r^{-2}$ dependence
for the medium density (Panaitescu \& Kumar 2000). In the optical/NIR
bands, it is predicted that, at the early epochs (up to 0.1 days after
explosion in the source rest frame), the temporal decay is significantly
flatter ($\alpha_1$ = 0.25) than found by us, while at later epochs our
measured $\alpha_2$ can be reproduced with $p$ = 2.1, which is
acceptable.

Then, we have made the hypothesis that the temporal break observed
in the optical light curves may be generated by the transition from
relativistic to non-relativistic conditions in the shock (Dai \& Lu
1999). By applying the prescriptions of Dai \& Lu (1999) to our
measured decay indices we obtain compatible conditions on $p$ only if the
$\nu_{\rm c}$ is above the optical band during our monitoring. From
$\alpha_1$ = 0.65$\pm$0.15 and $\alpha_2$ = 1.31$\pm$0.02 
we derive $p$ = 1.9$\pm$0.2 and $p$ = 2.27$\pm$0.01,
respectively, which we consider marginally consistent with 
$p \sim$ 2.1-2.2. This appears to be an acceptable value for the
electron distribution shape.
The predicted spectrum, for $\nu_{\rm opt} < \nu_{\rm c}$, is $\beta
\simeq$ 0.6, much flatter than observed. We have made the
hypothesis that intrinsic absorption at the source produces the
observed steeper spectrum and we have searched for a de-extinction
curve to correct for this.

As noted by Jha et al. (2001b), the optical spectrum lacks, at rest
frame $z$ = 1.475, the typical 2175 \AA~dust absorption wide
feature seen in the reddening curve of the Milky Way (Pei 1992,
Cardelli et al. 1989) and of star forming galaxies (e.g. Calzetti
1997). Thus, intrinsic absorption must not be large. We have
considered both an extinction curve appropriate for starburst
galaxies (Calzetti 1997) and the SMC curve (Pei 1992), and applied
them to our data. With the latter curve we cannot reproduce the
predicted spectral slope in the optical/NIR wavelength range
for any value of $E(B-V)$ (while in the optical range only,
acceptable fits are obtained, see also Lee et al. 2001). The
starburst curve provides instead an acceptable correction and
reproduces power-law spectra (except for the deviating $K$-band
point at epoch 0.97 days) with index $\beta \sim 0.5$ for a
moderate color excess, $E(B-V) \sim 0.35$ (see Fig. 6). The corrected
optical/NIR spectral index is much flatter than that found in X--rays,
which suggests that $\nu_{\rm c}$ lies between the optical and X--ray
bands. This is strengthened by the consistency of the X--ray spectral
index with $p/2$.

An empirical value of $\nu_{\rm c} \simeq 3.5 \times 10^{15}$ Hz
(supposed to be nearly constant during the non-relativistic expansion, Dai
\& Lu 1999) is determined through the extrapolation of the optical and
X--ray spectra (Fig. 6). The hypothesis of extinction within a heavily
star forming host galaxy has the advantage of removing the mismatch
between the optical and X--ray spectra normalizations (otherwise
attributable to other causes, like inverse Compton emission in the
X--rays, as suggested by in 't Zand et al. 2001), and had been applied in
a similar context also to GRB971214 (Dal Fiume et al. 2000).

The $J$-band light curve behaviour is similar to that of the
optical ones, indicating that the injection break frequency $\nu_{\rm m}$
is below this band at both epochs of NIR observations. However, the
behaviour of the $K$-band light curve deviates significantly from
that observed at shorter wavelengths. Comparison of the broad-band
spectra at 0.97 and 1.8 days after GRB could suggest that the
injection break may be located between the $K$ and $J$ bands at the
former epoch, and has moved below the $K$ band at the latter. 
Although this would be consistent with the evolution time scale of
$\nu_{\rm m}$ ($\propto t^{-3/2}$), the higher fluxes detected at sub-mm
wavelengths at preceding and successive epochs, indicate that
the injection break should be located longward of the NIR band during our
monitoring (Fich et al. 2001; Kulkarni et al. 2001). Therefore, we have no
straightforward explanation for the deviation of the $K$-band point from a
single optical-to-NIR power-law at 0.97 days.

At $z$ = 1.475, the total isotropic energy emitted in the 2-700 keV
range by this GRB is 7.7$\times$$10^{53}$ erg (in 't Zand et al.
2001).  With this total energy, and assuming a circumburst medium density
of $n=10^5-10^6$ cm$^{-3}$, the epoch of the transition from
ultrarelativistic to non-relativistic conditions is about 1 day (Dai \& Lu
1999). Therefore, the most viable interpretation of our data appears to
be a shock undergoing a substantial deceleration and transition to
non-relativistic conditions at about 0.5 days after the GRB (see also Dai
\& Cheng 2001). The consequent requirement of a thick ambient medium
(denser than $10^6$ cm$^{-3}$) is supported by the observation of the
absorbing system at $z$ = 1.475, likely coinciding with the redshift of
the source. This density would be typical for starburst galaxies (Calzetti
1997) or DLAs (Wolfe et al. 1986; Rao \& Turnshek 2000), and is similar to
that found for a number of GRB hosts (GRB990510, Vreeswijk et al.
2001; GRB000301C, Jensen et al. 2001; GRB000926, Fynbo et al.
2001a). Therefore, as for many previous OTs, there is the
suggestion that GRB010222 occurred in a site of substantial star
formation. 

Our observations and conclusions point to the importance of early
optical/NIR monitoring of GRB afterglows to establish their behaviour and
to get insight into the GRB progenitor. This will be guaranteed by the
current (HETE-2) or soon-to-fly GRB missions ({\it AGILE} and {\it
Swift}). 

\begin{acknowledgements}

We thank the staff astronomers of the TNG, Asiago, Loiano, Campo
Imperatore and TIRGO Observatories. We also thank the anonymous referee
for comments and suggestions which helped us improving the paper. We
acknowledge Scott Barthelmy for maintaining the GRB Coordinates Network
(GCN) and BACODINE services; the {\it BeppoSAX} e-mail GRB Alert Service
is also acknowledged. Krzysztof Stanek is thanked for having noted a
discrepancy between our $U$-band calibration and his, and for exchanging
his $U$-band data with us. C. Bartolini, A. Guarnieri, and A. Piccioni
acknowledge the University of Bologna (Funds for Selected Research
Topics).

\end{acknowledgements}


\begin{thebibliography}{}

\bibitem{} Bartolini, C., Bernabei, S., Guarnieri, A., et al., 2001, GCN
	982

\bibitem{} Berger, E., \& Frail, D.A., 2001, GCN 968

\bibitem{} Bersanelli, M., Bouchet, P., \& Falomo, R., 1991, A\&A, 252,
	854 

\bibitem{} Beuermann, K., Hessman, F.V., Reinsch, K., et al., 1999, A\&A,
	352, L26 

\bibitem{} Billings, G., 2001, GCN 969

\bibitem{} Bloom, J.S., Djorgovski, S.G., Halpern, J.P., et al., 2001, GCN
	989

\bibitem{} Calzetti, D., 1997, AJ, 113, 162

\bibitem{} Cardelli, J.A., Clayton, G.C., \& Mathis J.S., 1989, ApJ, 345,
	245

\bibitem{} Castro, S., Djorgovski, S.G., Kulkarni, S.R., et al., 2001, GCN
	999

\bibitem{} Castro-Tirado, A.J., Gorosabel, J., Zapatero-Osorio, M.R., et
	al., 1999, Science, 283, 2069

\bibitem{} Castro-Tirado, A.J., Sokolov, V.V., Gorosabel, J., et al.,
	2001, A\&A, 370, 398

\bibitem{} Chevalier, R.A., \& Li, Z.Y., 1999, ApJ, 520, L59

\bibitem{} Chevalier, R.A., \& Li, Z.Y., 2000, ApJ, 536, 195

\bibitem{} Cowsik, R., \& Bhargavi, S.G., 2001, GCN 1051

\bibitem{} Dai, Z.G., \& Lu, T., 1999, ApJ, 519, L155

\bibitem{} Dai, Z.G., \& Cheng, K.S., 2001, ApJ, submitted
	(astro-ph/0105055)

\bibitem{} Dal Fiume, D., Amati, L., Antonelli, L.A., et al., 2000, A\&A,
	355, 454 

\bibitem{} Di Paola, A., Antonelli, L.A., Li Causi, G., \& Valentini, G.,
	2001, GCN 977

\bibitem{} Fich, M., Phillips, R.M., Tilanus, R.P.J., Frail, D.A., \& 
	Smith, I., 2001, GCN 971

\bibitem{} Frontera, F., Amati, L., Costa, E., et al., 2000, ApJS, 127, 59

\bibitem{} Fruchter, A.S., Thorsett, S.E., Metzger, M.R., et al. 1999,
	ApJ, 519, L13

\bibitem{} Fukugita, M., Shimasaku, K., \& Ichikawa, T., 1995, PASP, 107,
	945

\bibitem{} Fynbo, J.P.U., Gorosabel, J., Dall, T.H., et al., 2001a, A\&A,
	in press (astro-ph/0102158)

\bibitem{} Fynbo, J.P.U., Gorosabel, J., Jensen, B.L., et al., 2001b, GCN
	975

\bibitem{} Gandolfi, G., 2001, GCN 966

\bibitem{} Garnavich, P.M., Pahre, M.A., Jha, S., et al., 2001, GCN 965

\bibitem{} Halpern, J.P., Uglesich, R., Mirabal, N., et al., 2000, ApJ,
	543, 697

\bibitem{} Harrison, F.A., Bloom, J.S., Frail, D.A., et al., 1999, ApJ,
	523, L121

\bibitem{} Henden, A., 2001a, GCN 961

\bibitem{} Henden, A., 2001b, GCN 962

\bibitem{} Henden, A., 2001c, GCN 987

\bibitem{} Henden, A., \& Vrba F., 2001, GCN 967

\bibitem{} Horne, K., 1986, PASP, 98, 609

\bibitem{} Hunt, L.K., Mannucci, F., Testi, L., et al., 1998, AJ, 115,
	2594

\bibitem{} in 't Zand, J.J.M., Kuiper, L., Amati, L., et al., 2001, ApJ,
	submitted, astro-ph/0104362

\bibitem{} Israel, G.L., Marconi, G., Covino, S., et al., 1999, A\&A, 348,
	L5

\bibitem{} Jaunsen, A.O., Hjorth, J., Bj\"ornsson, G., et al., 2001, ApJ,
	546, 127

\bibitem{} Jensen, B.L., Fynbo, J.P.U., Gorosabel, J., et al., 2001, A\&A,
	370, 909

\bibitem{} Jha, S., Matheson, T., Calkins, M., et al., 2001a, GCN 974

\bibitem{} Jha, S., Pahre, M.A., Garnavich, P.M., et al., 2001b, ApJ,
	in press, (astro-ph/0103081)

\bibitem{} Klose, S., Stecklum, B., Masetti, N., et al., 2000, ApJ, 545,
	271

\bibitem{} Kulkarni, S.R., Djorgovski, S.G., Odewahn S.C., et al., 1999,
	Nat, 398, 389

\bibitem{} Kulkarni, S.R., Frail, D.A., Moriarty-Schieven, G., et al.,
	2001, GCN 996

\bibitem{} Lee, B.C., Tucker, D.L., Vanden Berk, D.E., et al., 2001, ApJ,
	submitted (astro-ph/0104201)

\bibitem{} Masetti, N., Palazzi, E., Pian E., et al., 2000a, A\&A, 354,
	473

\bibitem{} Masetti, N., Bartolini, C., Bernabei, S., et al., 2000b, A\&A
	359, L23

\bibitem{} Masetti, N., Palazzi, E., Pian, E., et al., 2001, GCN 985

\bibitem{} Massey, P., Strobel, K., Barnes, J.V., \& Anderson, E., 1988,
	ApJ, 328, 315

\bibitem{} McDowell, J., Kilgard, R., Garnavich, P.M., Stanek, K.Z.,
	\& Jha, S., 2001, GCN 963

\bibitem{} Palazzi, E., Pian, E., Masetti N., et al., 1998, A\&A, 336, L95

\bibitem{} Panaitescu, A., \& Kumar, P., 2000, ApJ, 543, 66

\bibitem{} Pei, W., 1992, ApJ, 395, 130

\bibitem{} Piro, L., 2001a, GCN 959

\bibitem{} Piro, L., 2001b, GCN 960

\bibitem{} Price, P.A., Harrison, F.A., Galama, T.J., et al., 2001a, ApJ,
	549, L7

\bibitem{} Price, P.A., Gal-Yam, A., Ofek, E., et al., 2001b, GCN 973

\bibitem{} Rao, S.M., \& Turnshek, D.A., 2000, ApJS, 130, 1

\bibitem{} Rhoads, J.E., 1999, ApJ, 525, 737

\bibitem{} Rieke, G.H., \& Lebofsky, M.J., 1985, ApJ, 288, 618

\bibitem{} Sagar, R., Stalin, C.S., Bhattacharya, D., et al., 2001, Bull.
	Astron. Soc. India, submitted (astro-ph/0104249)

\bibitem{} Sari, R., Piran, T., \& Narayan, R., 1998, ApJ, 497, L17

\bibitem{} Sari, R., Piran, T., \& Halpern, J.P., 1999, ApJ, 519, L17

\bibitem{} Schlegel, D.J., Finkbeiner, D.P., \& Davis, M., 1998, ApJ, 500,
	525

\bibitem{} \v{S}imon, V., Hudec, R., Pizzichini, G., \& Masetti, N., 2001,
	A\&A, submitted

\bibitem{} Stanek, K.Z., Garnavich, P.M., Kaluzny, J., et al., 1999, ApJ,
	522, L39

\bibitem{} Stanek, K.Z., Jha, S., McDowell, J., et al., 2001a, GCN 970

\bibitem{} Stanek, K.Z., Garnavich, P., Jha, S., \& Pahre, M., 2001b, IAU
	Circ. 7586

\bibitem{} Stanek, K.Z., Garnavich, P.M., Jha S., et al., 2001c, ApJ,
	submitted (astro-ph/0104329)

\bibitem{} Stetson, P.B., 1987, PASP, 99, 191

\bibitem{} Vreeswijk, P.M., Fruchter, A.S., Kaper, L., et al., 2001, ApJ,
	546, 672

\bibitem{} Wainscoat, R.J., \& Cowie, L.L., 1992, AJ, 103, 332

\bibitem{} Wang, X., \& Loeb, A., 2000, ApJ, 535, 788

\bibitem{} Wolfe, A.M., Turnshek, D.A., Smith, H.E., \& Cohen, R.D., 1986,
	ApJS, 61, 249

\end{thebibliography}
\end{document}